\documentclass[aps,prd,showpacs,preprintnumbers,nofootinbib,floatfix,floats,groupedaddress,twocolumn]{revtex4}

\usepackage{bm}
\usepackage{latexsym}
\usepackage{dcolumn}
\usepackage{amsmath,amsfonts,amssymb}
\usepackage{graphicx,epsfig}

\def\eq#1{{Eq.~(\ref{#1})}}
\def\l{\left}
\def\r{\right}
\def\DM{\mathrm{d}}
\def\hatxi{\widehat{\xi}}
\def\hatn{\widehat{n}}
\def\tn{\textsl n}
\def\tt{\textsl t}
\def\rnn{R^{\tn {\small A}}_{~ \tn {\small A}}}
\def\rtt{R^{\tt {\small A}}_{~ \tt {\small A}}}
\def\rnna{R^{\tn {\small A}}_{~ \tn {\small B}}}
\def\rtta{R^{\tt {\small A}}_{~ \tt {\small B}}}

\reversemarginpar

\begin{document}
\title{Thermodynamic structure of Lanczos-Lovelock field equations from near-horizon symmetries}
 \author{Dawood Kothawala}
 \email{dawood@iucaa.ernet.in}
 \author{T.~Padmanabhan}
 \email{paddy@iucaa.ernet.in}
 \affiliation{IUCAA,
 Post Bag 4, Ganeshkhind, Pune - 411 007, India}

\date{\today}
\begin{abstract}
It is well known that, for a wide class of spacetimes with horizons, Einstein equations near the horizon can be written as a thermodynamic identity. It is also known that the Einstein tensor acquires a highly symmetric form near static, as well as stationary, horizons. We show that, for generic static spacetimes, this highly symmetric form of the Einstein tensor leads quite naturally and generically to the interpretation of the near-horizon field equations as a thermodynamic identity. We further extend this result to generic static spacetimes in Lanczos-Lovelock gravity, and show that the near-horizon field equations again represent a thermodynamic identity in all these models. These results confirm the conjecture that this thermodynamic perspective of gravity extends far beyond Einstein's theory.
\end{abstract} 

\maketitle
\vskip 0.5 in
\noindent
\maketitle
\section{Introduction}
It is a well established fact that Einstein field equations, near a horizon, can be written as a thermodynamic identity \cite{gravtherm, gravtherm2}; moreover, the result also extends to spherically symmetric horizons in Lanczos-Lovelock (LL) gravity \cite{aseem-lovelock}. This fact lends support to the point of view that gravity is a long wavelength, emergent phenomenon, and gravitational dynamics, at the macroscopic level, is therefore governed by relations which bear resemblance to the equations of thermodynamics \cite{sakharov,paddy-grf}. Since we do have operationally well defined notions such as entropy and temperature associated with a wide class of horizons in general relativity, it is natural to expect that the \textit{near horizon} behaviour of the field equations of gravity might actually be a statement of local thermodynamic equilibrium. 

Earlier demonstrations of the thermodynamic structure of gravitational field equations have involved certain assumptions like, for example, that of spherically symmetry, which is somewhat restrictive. The main aim of this paper is to provide a general proof, based on the near-horizon symmetries of the LL field equations, that near \textit{any static horizon}, the field equations can be written as $T {\DM} S - {\DM} E = P_{\perp} {\DM} V$, where the variations correspond to normal displacement of the horizon.

The paper is structured as follows: In the next section, we review the case of spherically symmetric spacetimes in Einstein gravity to stress the essential ideas involved. In section \ref{sec:setup}, we define the coordinate system which is suitable to describe the  general static spacetime, and also specify its properties and role in subsequent developments. In section \ref{sec:einstein}, we use the near-horizon symmetries of the Einstein tensor to show that the relevant field equations represent a thermodynamic identity; the result in section \ref{sec:eh-review} is then easily seen to be a special case of this more general result. In section \ref{sec:lovelock}, we extend the analysis of section \ref{sec:einstein} to LL lagrangians, and prove that, even in this case, the near horizon symmetries lead to a thermodynamic interpretation of the field equations. The key equation in this section is Eq.\;(\ref{eq:ll-tensor-near-hor}), which gives the near horizon structure of the LL tensor. Finally, in sections \ref{sec:comments} and \ref{sec:discussion}, we comment on certain relevant issues related to the physical interpretation of the result, and suggest a couple of possible generalizations. 

The metric signature is $(-, +, +, \ldots ,+)$, and all the fundamental constants such as $G, \hbar$ and $c$ have been set to unity (except when specified otherwise, in section \ref{sec:eh-review}).  Latin indices run from 0-3, whereas Greek indices run from 1-3; also, the capitalized Latin indices stand for the transverse coordinates. To simplify notation, we frequently use $\DM \Sigma$ to denote the volume element of the transverse $(D-2)$-surface; that is, $\DM \Sigma = \DM^{\tiny{D-2}}y \; \sqrt{\sigma}$ where $\sigma$ is the determinant of the metric in the space spanned by the $(D-2)$ transverse coordinates $y^A$.
\section{Spherically symmetric spacetimes: Revisited} \label{sec:eh-review}
Consider a static, spherically symmetric spacetime with a horizon, described by the metric:  
\begin{equation}
ds^2 = -f(r) c^2 dt^2 + \frac{1}{f(r)} dr^2 +r^2 d\Omega^2.
\label{eq:sph-symm-gen}
\end{equation}
We assume that this spacetime has a horizon at $r=a$, determined by $f(a)=0$ (we assume that $f(r)$ only has \textit{simple} zeroes), and the surface gravity $\kappa = f^\prime(a)/2$ is finite there. Periodicity in Euclidean time allows us to associate a temperature with the horizon as $k_BT=\hbar c\kappa/2\pi=\hbar c f'(a)/4\pi$. (Even for spacetimes with multiple horizons, this prescription is locally valid for each horizon surface.) The only non-trivial Einstein equation for this metric is $rf'(r)-(1-f)=(8\pi G/c^4) Pr^2\,$ (where $P$ is the radial pressure); when evaluated at $r=a$, this equation gives 
\begin{equation}
\frac{c^4}{G}\left[\frac{1}{2} f'(a)a - \frac{1}{2}\right] = 4\pi P a^2 
\label{reqa}
\end{equation} 
Multiplying \eq{reqa} by d$a$, and introducing a $\hbar$ factor \textit{by hand} into an otherwise classical equation, we can rewrite it as 
\begin{equation}
    \underbrace{\frac{{{\hbar}} cf'(a)}{4\pi}}_{\displaystyle{k_BT}}
     \ \underbrace{\frac{c^3}{G{{\hbar}}}{{\DM}}\left( \frac{1}{4} 4\pi a^2
     \right)}_{ 
     \displaystyle{{{\DM}}S}}
   \ \underbrace{-\ \frac{1}{2}\frac{c^4 {{\DM}}a}{G}}_{
     \displaystyle{-{{\DM}}E}}
  = \underbrace{P {{\DM}} \left( \frac{4\pi}{3}  a^3 \right)  }_{
     \displaystyle{P\, {{\DM}}V}}
 \label{EHthermo}
\end{equation}
and read off the expressions:
\begin{equation*}
 S=\frac{1}{4L_P^2} (4\pi a^2) = \frac{1}{4} {\frac{A_H}{L_P^2}}; \quad
    E={\frac{c^4}{2G}} a  
    =\frac{c^4}{G}\left( \frac{A_H}{16\pi}\right)^{1/2}
\end{equation*} 
where $A_H$ is the horizon area and $L_P^2=G\hbar/c^3$. Thus, we see that, \textit{Einstein equations can be written simply as a thermodynamic identity}. 

Before proceeding further, we will make couple of comments regarding this result, which are relevant and shall remain valid for our generalization discussed in the rest of the paper: 
First,  the combination $T$d$S$ is completely \textit{classical}, and is independent of $\hbar$; however, $T \propto \hbar$ and $S \propto 1/\hbar$. This is analogous to the situation in classical thermodynamics when compared to statistical mechanics. The $T$d$S$ in thermodynamics is independent of the Boltzmann's constant while statistical mechanics will lead to an $S\propto k_B$  and $T\propto1/k_B$. But since Euclidean periodicity allows us to determine $T$ independently, we can immediately read-off $S$.

Second, in spite of the superficial similarity, \eq{EHthermo} is \textit{different} from the conventional first law of black hole thermodynamics,  due to the presence of the $P$d$V$ term. This relation is more in tune with the membrane paradigm \cite{membrane} for black holes. The difference is easily seen, for example, in the case of Reissner-Nordstrom black hole for which $P \neq 0$. If a \textit{chargeless} particle of mass d$M$ is dropped into a Reissner-Nordstrom blackhole, then an elementary calculation shows that the energy, defined above as $E = a/2$, changes by d$E= ({{\DM}}a/2)=(1/2)[a/(a-M)]{{\DM}}M \neq {{\DM}}M$ while it is ${{\DM}}E+P{{\DM}}V$ which is precisely equal to ${{\DM}}M$, making sure $T{{\DM}}S={{\DM}}M$. So we need the $P{{\DM}}V$ term to get $T{{\DM}}S={{\DM}}M$ when a chargeless particle is dropped into a Reissner-Nordstrom blackhole. More generally, if ${{\DM}}a$ arises due to changes ${{\DM}}M$ and ${{\DM}}Q$, it is easy to show that \eq{EHthermo} gives $T{{\DM}}S={{\DM}}M -(Q/a){{\DM}}Q$ where the second term arises from the electrostatic contribution from the horizon surface charge as expected in the membrane paradigm.

Our aim in the rest of the paper is to provide a general proof of this relation between gravitational field equations and horizon thermodynamics. We show that the thermodynamic structure arises essentially because of the \textit{near-horizon symmetries} of the gravitational field equations for general static spacetimes in not only Einstein gravity, but even in Lanczos-Lovelock theory. As we shall see, the generalization involves attaching specific meaning to the variations in the first law, and would lead to the above result (obtained for the spherically symmetric case) as a special case. In order to set the stage for the subsequent analysis, we now rewrite the above result for the spherically symmetric case in a slightly different manner.

As mentioned above, for the spacetimes described by metric (\ref{eq:sph-symm-gen}), we have
\begin{eqnarray*}
G^t_t = G^r_r = \frac{rf'-(1-f)}{r^2}
\end{eqnarray*}
We can rewrite the above expression using (i) the transverse metric, $\DM \Sigma=\sqrt{\sigma} \DM^2 y= r^2 \sin{\theta} \DM \theta \DM \phi$, (ii) the Ricciscalar, $R_{\parallel} = 2 / r^2$ calculated from the transverse metric $\sigma$, and (iii) the field equations to set $G^r_r = 8 \pi T^r_r$. Note that, since $R^{\theta \phi}_{\theta \phi} = (1 - f) / r^2$, we have, near the horizon, $R_{\parallel} = R^{\small{A B}}_{\small{A B}} + \mathcal{O}(r-a)$ (where we would like to remind of our notation in which capitalized Latin indices stand for the $(D-2)$ transverse coordinates). The Einstein equation on the horizon is therefore expressible in the form
\begin{eqnarray}
\frac{f'(a)}{4 \pi} \frac{\sqrt{\sigma}}{2 r} - \frac{1}{4} \l( \frac{1}{4 \pi} R_{\parallel} \sqrt{\sigma} \r) &=& T^r_r \sqrt{\sigma}
\end{eqnarray}
Further, since
\begin{equation}
\frac{\sqrt{\sigma}}{r} = \frac{1}{2} \frac{\partial}{\partial r} \sqrt{\sigma}
\end{equation}
we have,
\begin{eqnarray}
T \frac{\partial}{\partial r} \l( \frac{1}{4} \sqrt{\sigma} \r) - \frac{1}{4} \l( \frac{1}{4 \pi} R_{\parallel} \sqrt{\sigma} \r) &=& T^r_r \sqrt{\sigma}
\end{eqnarray}
Upon multiplying the above expression by $\delta r \; \DM \theta \DM \phi$, and integrating over the horizon 2-surface, we immediately obtain
\begin{eqnarray}
T \frac{\partial}{\partial r} \l[ \int \frac{1}{4} \DM \Sigma \r]_H {\delta r} &-& \l[ \int_H \frac{1}{8 \pi} R_{\parallel} \DM \Sigma \r]~ \frac{\delta r}{2} \nonumber \\
\nonumber \\
&=& \int_H P_r \DM \Sigma \; \delta r
\label{eq:connect-sph-symm-1}
\end{eqnarray}
This is the relation we wanted to establish; as we shall see, Eq.\;(\ref{eq:connect-sph-symm-1}) turns out to be 
quite general, and holds in arbitrary static spacetimes (in four dimensions) with the replacement of $r$ with the affine parameter along the outgoing null geodesics [see Eq.\;(\ref{eq:connect-sph-symm})]. 

Of course, the on-horizon limit in the first term is to be taken \textit{after} evaluating the derivative with respect to $r$. Formally, we can define a function $S$ of $r$ as 
\begin{eqnarray*}
S(r) = \int \frac{1}{4} \DM \Sigma \; ,
\end{eqnarray*}
the integral being over a $r=$constant surface. The derivative of this function at $r=a$ is well defined and finite, while the value of the function itself at $r=a$ is the Bekenstein-Hawking entropy of the black hole. 

For the spherically symmetric case we are dealing with, the angular integrations are trivial. \textit{In fact, the integral in the second term on the LHS of Eq.\;(\ref{eq:connect-sph-symm-1}) gives unity!} In general, this integral is one-half the Euler characteristic $\chi$ of the horizon 2-surface. When the horizon is a two-sphere, $\chi=2$, and the integral is unity.
\section{Einstein equations as a thermodynamic identity} \label{sec:einstein}
\subsection{Background} \label{sec:setup}
We shall now set up the coordinate system best suited for the discussion of the  general static spacetime, and identify the affine parameter for the outgoing null geodesics near the horizon. We begin with the metric \cite{visser1}
\begin{eqnarray}
{\DM}s^2 = - N^2 {\DM}t^2 + {\DM}n^2 + \sigma_{A B} {\DM}y^A {\DM}y^B
\label{static-metric}
\end{eqnarray}
where $\sigma_{A B}(n, y^A)$ is the transverse metric, and the Killing horizon, generated by the timelike Killing vector field $\bm{\xi} = \bm{\partial}_t$, is approached as $N^2 \rightarrow 0$. Near the horizon, $N \sim \kappa n + \mathcal{O}(n^3)$ where $\kappa$ is the surface gravity. The $t=$ constant part of the metric is written by employing \textit{Gaussian normal coordinates} for the spatial part of the  metric spanned by $\left( {n, y^A} \right)$, $n$ being the normal distance to the horizon. To determine the null geodesics for this spacetime, we rewrite the above metric as
\begin{eqnarray}
{\DM}s^2 = - N^2 \left( {\DM}t - N^{-1} {\DM}n \right) \left( {\DM}t + N^{-1} {\DM}n \right) + \sigma_{A B} {\DM}y^A {\DM}y^B
\nonumber \\
\end{eqnarray}
The $y^A = $ const. null geodesics are then given by 
\begin{eqnarray}
u &=& t - \int N^{-1} {\DM}n = \mathrm{constant} \nonumber
\\
v &=& t + \int N^{-1} {\DM}n = \mathrm{constant}
\end{eqnarray}
The tangent vector of outgoing and ingoing null geodesics are then 
\begin{eqnarray}
\bm{l} &=& - {\nabla} u = \left( -1, + N^{-1} \right)  \nonumber  
\\
\bm{k} &=& - {\nabla} v = \left( -1, - N^{-1} \right)
\label{null-geod}
\end{eqnarray}
As is easy to check, these vectors satisfy the geodesic equation in affinely parametrized form, that is; $\nabla_{\bm{l}} \bm{l} = 0 = \nabla_{\bm{k}} \bm{k}$. The affine parameter $\lambda$, defined by 
\begin{eqnarray}
\bm{l} \cdot \nabla \lambda = 1 \label{affine-condition}
\end{eqnarray}
can be found by noting that, near the horizon, $N \sim \kappa \; n$. Using this, we find that,
\begin{eqnarray}
\lambda \sim \lambda_H + \frac{1}{2} \; \kappa \; n^2  
\end{eqnarray}
where $\lambda = \lambda_H$ is the location of the horizon. [For $\bm{k}$, the affine parameter would be $\lambda_H - (1/2) \kappa n^2$.] Note that, $N^2 \bm{l} \rightarrow \bm{\xi}|_{H}$, which implies, $2 \kappa \l( \lambda - \lambda_H \r) \bm{l} \rightarrow \bm{\xi}|_{H}$. In subsequent analysis, the differentials of various geometric quantities (such as entropy) defined on the horizon, which are directly involved in the statement of the first law of thermodynamics, are to be interpreted as variations with respect to the affine parameter along the outgoing null geodesics, i.e., $\lambda$. This, of course, is the most natural variation that can be chosen on a \textit{null surface}.
\subsection{Near-horizon behaviour of Einstein tensor} \label{sec:einstein}
We will now use the near horizon symmetries of the Einstein tensor to prove that the field equations near the horizon have a thermodynamic interpretation. We begin with the following expression for the on-horizon structure of the Einstein tensor (which is derived in \cite{visser1}; for the sake of completeness, we give a proof in Appendix \ref{app1}):
\begin{eqnarray}
G^{\hatxi}_{\hatxi} \;|_H = G^{\hatn}_{\hatn} \;|_H = \frac{1}{2} \mathrm{tr} \left[ \sigma_2 \right] - \frac{1}{2} R_{\parallel} \label{near-hor-eins}
\end{eqnarray}
where $R_{\parallel}$ is the Ricci scalar of the on-horizon transverse metric, $\left[ \sigma_H \right]_{A B}$, and $\sigma_2$ is defined by 
\begin{eqnarray}
\sigma_{A B} &=& \left[ \sigma_H(y) \right]_{A B} + \frac{1}{2} \left[ \sigma_2(y) \right]_{A B} n^2 + \mathcal{O}(n^3) \nonumber 
\nonumber \\
\nonumber \\
&=& \left[ \sigma_H (y; \lambda_H) \right]_{A B} + {\kappa}^{-1} \left[ \sigma_2(y; \lambda_H) \right]_{A B} \l( \lambda - \lambda_H \r) 
\nonumber 
\\ 
\nonumber 
\\ 
&& \hspace{1.0in} + \mathcal{O}( (\lambda - \lambda_H)^{3/2} ) \nonumber
\\
\label{sigma-taylor}
\end{eqnarray}
where the second expression makes it clear that the on-horizon transverse metric will --- in general --- depend on the \textit{parameter} $\lambda_H$. (In spherical symmetry, this is in fact the only dependence of the on-horizon transverse metric on parameters such as mass, charge etc.; however, this will not be true in general, as in the case of Kerr spacetime where there is an additional, explicit dependence on the rotation parameter.)  The absence of a term linear in $n$ in the above expansion follows from the requirement that the curvature invariants be finite on the horizon \cite{visser1}. The Einstein tensor components given above are evaluated in an orthonormal tetrad appropriate for a timelike observer moving along the orbit of the Killing vector field generating the Killing horizon. This is denoted by a hat on the indices; for example, $\bm{\hatxi} = \left( - g_{t t} \right)^{-1/2}~\bm{\partial}_t$ etc., and $-G^{\hatxi}_{\hatxi} = G_{{\hatxi} {\hatxi}} = G( \bm{\hatxi}, \bm{\hatxi} )$. Also, the trace operation, $\mathrm{tr}$, is performed using the transverse metric $\sigma_{A B}$, but can as well be performed using $\left[ \sigma_H \right]_{A B}$ in the $n \rightarrow 0$ limit. The validity of expression (\ref{near-hor-eins}), with the given Taylor series expansions for $N(n, y^{\small A})$ and $\sigma_{A B}(n, y^{\small A})$, can be easily checked using a symbolic package such as \texttt{MAPLE}. \footnote{While using \texttt{MAPLE}, it is easier to verify Eq.~(\ref{near-hor-eins}) by first asking \texttt{MAPLE} to evaluate $R_{\parallel}$ for a general 2-D metric, add it to the components of the on-horizon Einstein tensor, and then take the limit $n \rightarrow 0$; although all the intermediate expressions appear awful, the remainder is easily seen to be $\mathrm{tr} \left[ \sigma_2 \right] / 2$.}

We will express $G^{\hatxi}_{\hatxi}$ in terms of the variation of the transverse area with respect to the affine parameter. To do this,  consider the variation of the transverse area in the \textit{normal} direction, with respect to the affine parameter $\lambda$,
\begin{eqnarray}
\delta_{\lambda} \sqrt{\sigma} &=& l^c \partial_c \sqrt{\sigma} ~ \delta \lambda \nonumber \\
&=& \left( \frac{\partial}{\partial \lambda} \sqrt{\sigma} \right) ~ \delta \lambda \nonumber \\ 
&=& \frac{1}{2} \sqrt{\sigma} ~ \sigma^{A B} \left( \frac{\partial}{\partial \lambda} \; \sigma_{A B} \right) ~ \delta \lambda 
 \nonumber \\
&=& \frac{1}{2 \kappa} \sqrt{\sigma} ~ \mathrm{tr} \left[ \sigma_2 \right] ~ \delta \lambda
\end{eqnarray}
Therefore, we obtain, on multiplying Eq.\;(\ref{near-hor-eins}) by $\delta \lambda$,
\begin{eqnarray}
G^{\hatxi}_{\hatxi} \; \delta \lambda = G^{\hatn}_{\hatn} \; \delta \lambda = \kappa \; \frac{\delta_{\lambda} \sqrt{\sigma}}{\sqrt{\sigma}} - \frac{1}{2} R_{\parallel} \; \delta \lambda
\end{eqnarray}
Re-arranging this expression, we get
\begin{eqnarray}
\frac{\kappa}{2 \pi} \frac{\partial}{\partial \lambda} \left( \frac{1}{4} \sqrt{\sigma} \right) {\delta \lambda} - \left\{ \frac{1}{8 \pi} R_{\parallel} \sqrt{\sigma} \right\} \; \frac{\delta \lambda}{2} &=& \frac{1}{8 \pi} G^{\hatxi}_{\hatxi} \sqrt{\sigma} \; \delta \lambda 
 \nonumber \\
 \nonumber \\
&=& \frac{1}{8 \pi} G^{\hatn}_{\hatn} \sqrt{\sigma} \; \delta \lambda
 \nonumber \\
 \nonumber \\
&=& T^{\hatn}_{\hatn} \sqrt{\sigma} \; \delta \lambda  \nonumber \\
\end{eqnarray}
where we have used $G^{\hatxi}_{\hatxi} \;|_H = G^{\hatn}_{\hatn} \;|_H$ in the second line and Einstein equation in the third line. Upon multiplying the above expression by ${\DM}^2 y$, and integrating over the horizon 2-surface, we immediately obtain
\begin{eqnarray}
T \frac{\partial}{\partial \lambda} \l[ \int \frac{1}{4} \sqrt{\sigma} \; {\DM}^2 y \r]_H {\delta \lambda} &-& \l[ \int_H \frac{1}{8 \pi} R_{\parallel} \sqrt{\sigma} \; {\DM}^2 y \r]~ \frac{\delta \lambda}{2} \nonumber \\
\nonumber \\
&=& \int_H P_{\perp} \sqrt{\sigma} \; {\DM}^2 y \; \delta \lambda
\label{eq:connect-sph-symm}
\end{eqnarray}
where we have identified $ T = \kappa / 2 \pi $ as the horizon temperature, and used the interpretation of $T^{\hatn}_{\hatn}$ as normal pressure, $P_{\perp}$, \textit{on the horizon}. We can therefore interpret 
\begin{equation}
\overline{F} = \int_H P_{\perp} \sqrt{\sigma} \; {\DM}^2 y 
\end{equation} 
as the the average normal force over the horizon ``surface" (in the spirit of membrane paradigm) and $\overline{F} ~ {\delta} \lambda$ as the virtual work done in displacing the horizon by an affine distance ${\delta} \lambda$. The above equation can now be written as 
\begin{eqnarray}
T {\delta}_{\lambda} S - {\delta}_{\lambda} E = \overline{F} ~ {\delta} \lambda \label{first-law-2}
\end{eqnarray}
where 
\begin{eqnarray}
S = \frac{1}{4} \int \sqrt{\sigma} \; {\DM}^2 y
\end{eqnarray}
is (a priori) just a function of $\lambda$; in particular, the derivative of $S$ with respect to $\lambda$ is well-defined and finite on the horizon. We only need the expression for $S$ very close to the horizon. The value of $S$ at $\lambda=\lambda_H$,
\begin{eqnarray}
S\l( \lambda=\lambda_H \r) = \frac{1}{4} \int_H \sqrt{\sigma} \; {\DM}^2 y
\end{eqnarray}
is equal to the Bekenstein-Hawking entropy of the horizon. (We do not attribute any physical significance  to the value of $S$ away from the horizon.) We have also identified the energy $E$ associated with the horizon as
\begin{eqnarray}
E = \l( \frac{\chi}{2} \r) \frac{\lambda_H}{2} \label{energy}
\end{eqnarray}
where $\chi$ is the Euler characteristic of a 2-dimensional compact \footnote{If the manifold has a boundary, then the expression for Euler characteristic will have additional boundary terms.} manifold $\mathcal{M}_2$ [which in this case would be the horizon 2-surface], given by
\begin{eqnarray}
\chi \left( \mathcal{M}_2 \right) = \frac{1}{4 \pi} \int_{\mathcal{M}_2} ~ R ~ \mathrm{d[vol]}
\end{eqnarray}
Some comments are in order regarding the expression for $E$:
First, we have set the arbitrary integration constant in Eq.~(\ref{energy}) to zero. With this choice, we have chosen the affine parameter such that $E \rightarrow 0$ as $\lambda_H \rightarrow 0$; that is, if one considers a \textit{class} of static spacetimes parametrized by $\lambda_H$, then our choice implies that $E$ vanishes when $\lambda_H = 0$ (In the simple context of Schwarschild metric, for example, we have $\lambda_H=2M$ and this condition says that the energy vanishes when $M=0$). In general, any non-trivial solution of the field equations with a horizon will depend on several parameters, say $\{ \alpha_i \}$ (for e.g., the mass $M$ and charge $Q$ in case of the Reissner-Nordstrom  solution). The parameter $\lambda_H$ which fixes the horizon location will be a function of these parameters, i.e., $\lambda_H = \lambda_H(\{\alpha_i\}$. Our choice for $E$ is therefore equivalent to demanding that $\lambda_H(\{0\})=0$ so that $E$ goes to zero when there is no horizon.  This fixes the choice of the additive constant in the affine parameter which we mentioned before.
 
 Second point, which is more important, is that our particular identification of $E$ is fixed by the choice of the affine parameter along the outgoing null geodesics (see Eqs.~(\ref{null-geod})). In particular, this brings out the significance of the \textit{radial} coordinate $r$ in spherically symmetric and stationary spacetimes; in either case, $r$ is the affine parameter along the outgoing null geodesics. 

As is evident, these comments will also apply to the LL case discussed in the next section. To clarify these points further, let us consider the spherically symmetric case with a compact horizon, where $\lambda=r$, and $\chi=2$. We obtain $E=r_H/2$, $r_H$ being the horizon radius, which matches with the standard expression for quasilocal energy for such spacetimes obtained previously. In general, for a compact, simply connected horizon 2-surface, $\chi=2$ (since any such manifold is homeomorphic to a 2-sphere), and we have, $E = \lambda_H / 2$. Therefore, for spherically symmetric black holes, since $P_{\perp}=P_r$ is independent of the transverse coordinates $\left( \theta, \phi \right)$, we obtain
\begin{eqnarray}
T \delta S - \delta E = P_r \delta V
\end{eqnarray}
where, now, $\delta S = 2 \pi r_H \delta r_H$, $\delta E = \delta r_H / 2$, $P_r = T^r_r(r)|_{r=r_H}$, $\delta V = 4 \pi r_H^2 ~ \delta r_H$ and $T$ is the standard Hawking temperature. We therefore recover the result mentioned in section \ref{sec:eh-review}. As we shall see, exactly similar structure emerges for the near-horizon field equations of LL gravity as well.  
\section{Generalization to Lanczos-Lovelock gravity} \label{sec:lovelock}
We now turn attention to the more general case of LL lagrangians, and show that the near horizon structure of the field equations represent a thermodynamic identity even in this case.

The $m^{\mathrm{th}}$ order LL lagrangian in $D-$dimensions is given by (we will use the notations of \cite{aseem-lovelock} throughout this section)
\begin{eqnarray}
\mathcal{L}_m^{(D)} = \frac{1}{16 \pi} \frac{1}{2^m} \delta^{a_1 b_1 \ldots a_m b_m}_{c_1 d_1 \ldots c_m d_m} R^{c_1 d_1}_{~ a_1 b_1} \cdots R^{c_m d_m}_{~ a_m b_m}
\end{eqnarray}
A general LL action is given by linear combination of $\mathcal{L}_m^{(D)}$ for different $m$'s, with arbitrary constant coefficients, say $c_m$'s. The equations of motion are given by
\begin{eqnarray}
E^a_{b} &=& \sum_{m} {c_m E^{a}_{b(m)}}=\frac{1}{2} \; T^a_{b}\,,
\nonumber
\end{eqnarray}
where
\begin{eqnarray}
E^i_{j(m)} &=&  - \frac{1}{2} \frac{1}{16 \pi} \frac{1}{2^m} \delta^{i a_1 b_1 \ldots a_m b_m}_{j c_1 d_1 \ldots c_m d_m} R^{c_1 d_1}_{~ a_1 b_1} \cdots R^{c_m d_m}_{~ a_m b_m}
\nonumber \\
&=& \frac{1}{16 \pi} \frac{m}{2^m} \delta^{a_1 b_1 \ldots a_m b_m}_{j_{~} d_1 \ldots c_m d_m} R^{i d_1}_{~ a_1 b_1} \cdots R^{c_m d_m}_{~ a_m b_m} - \frac{1}{2} \delta^i_{j} \mathcal{L}_m
\nonumber \\
\label{eom-lovelock}
\end{eqnarray}
The equivalence of the two expressions given above can be easily established, see Appendix \ref{app:LLeom}.
We shall prove our result for a given $m$, and drop the subscript on $E^i_{j(m)}$ henceforth; the result for any linear combination follows immediately.  

It is possible to analyze the near horizon symmetries of the equations of motion using the Taylor series for $N(n, y^{\small A})$ and $\sigma_{\small{A B}}(n, y^{\small A})$ (which depend only on the finiteness of the curvature invariants as $n \rightarrow$ 0); see Appendix \ref{app2} for a brief discussion on this. 
We begin with the $t-t$ component of the field equations, and, following an analysis similar to the one leading to Eq.~(30) of \cite{aseem-lovelock}, we arrive at,
\begin{eqnarray}
E^{\hatxi}_{\hatxi} &=& E^\tt_\tt = \frac{1}{16 \pi} \frac{m}{2^m} \left[ \sigma_2 \right]_{C B}  \sigma^{C A} \mathcal{E}^B_A - \frac{1}{2} \mathcal{L}^{(D-2)}_m + \mathcal{O}(n) 
\nonumber \\
\end{eqnarray}
where
\begin{eqnarray}
\mathcal{E}^B_A &=& \delta^{B A_1 \ldots B_{m-1}}_{A C_1 \ldots D_{m-1}} ~ ^{(D-2)}R^{C_1 D_1}_{~ A_1 B_1} \cdots ~ ^{(D-2)}R^{C_{m-1} D_{m-1}}_{~ A_{m-1} B_{m-1}}  \nonumber \\ 
\end{eqnarray}
In Appendix \ref{app2}, we show that $E^{\hat \xi}_{\hat \xi} = E^{\hatn}_{\hatn}$ on the horizon. Therefore, we have
\begin{eqnarray}
E^{\hatxi}_{\hatxi} \;|_H = E^{\hatn}_{\hatn} \;|_H = \frac{1}{16 \pi} \frac{m}{2^m} \left[ \sigma_2 \right]_{C B}  \sigma^{C A} \mathcal{E}^B_A - \frac{1}{2} \mathcal{L}^{(D-2)}_m \nonumber 
\\ \label{eq:ll-tensor-near-hor}
\end{eqnarray}
which generalizes Eq.~(\ref{near-hor-eins}) to general LL lagrangians. We now again use the Taylor series expansion (\ref{sigma-taylor}), and the affine parameter $\lambda$, to obtain
\begin{eqnarray}
\delta_{\lambda} \sigma_{A B} = \frac{\delta \lambda}{\kappa} \left[ \sigma_2 \right]_{A B} + \mathcal{O}[ (\lambda - \lambda_H)^{1/2} \; \delta \lambda ]
\end{eqnarray}
Using this, we obtain
\begin{eqnarray}
2 E^{\hatxi}_{\hatxi} \; \sqrt{\sigma} \; \delta \lambda &=& T \l( \frac{1}{8} \frac{m}{2^{m-1}} \r) \mathcal{E}^{B C} \; \delta_{\lambda} \sigma_{B C} \; \sqrt{\sigma} 
\nonumber \\
\nonumber \\
&-& \mathcal{L}^{(D-2)}_m \sqrt{\sigma} \; \delta \lambda + \mathcal{O}[ (\lambda - \lambda_H)^{1/2} \; \delta \lambda] \nonumber 
\\
\label{eom-tds-1}
\end{eqnarray}
where we have introduced the Hawking temperature, $T$, in the last expression. We can now show that the factor multiplying $T$ is directly related to the variation of the following quantity, the variation being evaluated at $\lambda=\lambda_H$:
\begin{eqnarray}
S = 4 \pi m \int \DM \Sigma \; \mathcal{L}^{(D-2)}_{m-1} \label{wald-entropy-2} 
\end{eqnarray}
We simply note that the variation of the above expression must give equations of motion for the $(m-1)^{\mathrm{th}}$ order LL term in $(D-2)$ dimensions. (The variation would also produce surface terms, which would not contribute when evaluated at $\lambda=\lambda_H$ because the horizon is a compact surface with no boundary.) We have:
\begin{eqnarray}
\delta_{\lambda} S = - 4 \pi m \int_H \DM \Sigma \; \mathcal{E'}^{B C} ~ \delta_{\lambda} \sigma_{B C} 
\end{eqnarray}
where we have evaluated the variation on $\lambda=\lambda_H$. Noting that the lagrangian is
\begin{eqnarray*}
\mathcal{L}^{(D-2)}_{m-1} = \frac{1}{16 \pi} \frac{1}{2^{(m-1)}} \delta^{A_1 B_1 \ldots B_{m-1}}_{C_1 D_1 \ldots D_{m-1}} ~ \cdots ~ ^{(D-2)}R^{C_{m-1} D_{m-1}}_{~ A_{m-1} B_{m-1}} 
\end{eqnarray*}
and using the first of Eqs.~(\ref{eom-lovelock}), we see that
\begin{eqnarray}
\mathcal{E'}^B_C = - \frac{1}{2} \frac{1}{16 \pi} \frac{1}{2^{(m-1)}} \mathcal{E}^B_C
\end{eqnarray}
Therefore, we obtain
\begin{eqnarray}
\delta_{\lambda} S = \frac{1}{8} \frac{m}{2^{(m-1)}} \int_H \DM \Sigma \;\; \mathcal{E}^{B C} ~ \delta_{\lambda} \sigma_{B C}
\end{eqnarray}
\textit{which is precisely the integral of the factor multiplying $T$ in Eq.\;(\ref{eom-tds-1})}. As mentioned above, $S$ defined in Eq.\;(\ref{wald-entropy-2}) is a function of $\lambda$, and its derivative with respect to $\lambda$ is well defined and finite on the horizon. The expression for $S$, evaluated at $\lambda=\lambda_H$, 
\begin{eqnarray}
S\l( \lambda=\lambda_H \r) = 4 \pi m \int_H \DM \Sigma \;\; \mathcal{L}^{(D-2)}_{m-1} \label{wald-entropy-1} 
\end{eqnarray}
is what we shall interpret as the entropy of the horizon. (As mentioned before, $S$ has no physical significance away from the horizon.) \footnote{As an aside, we would also like to point out the following fact; when $D=2m$, that is, when the corresponding LL action is the Euler characteristic of the full $D$-dimensional manifold, then $D-2=2(m-1)$, and we see from Eq.~(\ref{wald-entropy-1}) that $S$ is proportional to the Euler characteristic of the $(D-2)$-dimensional horizon surface, determined by $\mathcal{L}^{(D-2)}_{m-1}$; therefore, $S$ is just a constant number. For example, the Gauss-Bonnet term ($m=2$) in $D=4$ will contribute a constant additive term to the standard Bekenstein-Hawking entropy.}  

Multiplying Eq.~(\ref{eom-tds-1}) by ${\DM}^{(D-2)} y$, integrating over the horizon surface, and taking the $n \rightarrow 0$ limit, we now see that it can be written as
\begin{eqnarray}
T {\delta}_{\lambda} S - \int_H \DM \Sigma \;\; \mathcal{L}^{(D-2)}_{m} ~ {\delta} \lambda  &=& \int_H \DM \Sigma \;\; T^{\hatxi}_{\hatxi} ~ {\delta} \lambda
\nonumber \\
&=& \int_H \DM \Sigma \;\; T^{\hatn}_{\hatn} ~ {\delta} \lambda
\nonumber \\
&=& \int_H \DM \Sigma \;\; P_{\perp} ~ {\delta} \lambda
\end{eqnarray}
where we have used the field equations $E^{\hatxi}_{\hatxi} = (1/2) T^{\hatxi}_{\hatxi}$ in the first equality, and the relation $E^{\hatxi}_{\hatxi} \;|_H = E^{\hatn}_{\hatn} \;|_H$ in the second equality. This equation now has the desired form of the first law of thermodynamics, provided: (i) We identify  the quantity $S$, defined by Eq.~(\ref{wald-entropy-1}) as the entropy of horizons in LL gravity; indeed, \textit{exactly} the same expression for entropy has been obtained in the literature using independent methods, see e.g, ref.\cite{entropyLL}. (ii) We also identify the second term on LHS as $\delta_{\lambda} E$; this leads to the definition of $E$ to be
\begin{eqnarray}
E = \int^{\lambda} \delta \lambda \int_H \DM \Sigma \;\; \mathcal{L}^{(D-2)}_{m}
\label{energy-ll}
\end{eqnarray}
where the $\lambda \rightarrow \lambda_H$ limit must be taken \textit{after} the integral is done (therefore, we need to know the detailed form of $\mathcal{L}^{(D-2)}_{m}$ as a function of $\lambda$ to calculate this explicitly). For $D = 2 (m+1)$, the integral over $H$ above is related to the Euler characteristic of the horizon, in which case $E \propto \lambda_H$ (where we have set the arbitrary integration constant to zero). For $m=2, D=4$, this reduces to the expression obtained earlier in the case of Einstein gravity. 

Let us briefly comment on the general form of $E$ for \textit{spherically symmetric} spacetimes for general LL lagrangians, with horizon at $r=r_H$, and $\lambda=r$. In this case, $\mathcal{L}^{(D-2)}_{m} \sim (1/{\lambda}^2)^{m}$ and $\sqrt{\sigma} \sim {\lambda}^{D-2}$. The integrand therefore scales as ${\lambda}^{(D-2) - 2 m}$. Integrating the RHS of Eq.~(\ref{energy-ll}), and taking the $\lambda \rightarrow \lambda_H$ limit after integration, we see that $E \sim \lambda_H^{(D-2) - 2 m + 1}$. As mentioned above, for $D = 2 (m+1)$, $E \sim \lambda_H$. In fact, in the case of spherically symmetric spacetimes in LL theory, the above expression can be formally shown to be \textit{exactly equivalent} to the one derived by others (see \cite{aseem-lovelock}, and also Ref.~[16] therein). 

As far as we are aware, no general expression for energy in LL theory exists in the literature, and ours could be thought of as first such definition which appears to be reasonable from physical point of view. The expression clearly deserves further investigation.

Putting all this together, we see that, for \textit{generic static spacetimes} in LL gravity, the field equations can be written as a thermodynamic identity:
\begin{eqnarray}
T \delta_{\lambda} S - \delta_{\lambda} E = \overline{F} \delta \lambda  
\end{eqnarray} 
thereby showing that the thermodynamic relations are far more general than Einstein field equations.
\section{Some comments on the result} \label{sec:comments}
It must be emphasized that, in the above derivation, we had no choice whatsoever in the expressions for $S$ and $E$. Once we have identified the work term, $\overline{F} \delta \lambda$, we \textit{must} choose the factor multiplying $T$ on the left hand side as $\delta_{\lambda} S$, and the remaining term as $-\delta_{\lambda} E$. Note that, the near-horizon structure implies $P_{\perp} \sqrt{\sigma} \delta \lambda = - T^{a}_{b} \xi^b l_a \sqrt{\sigma} \; \delta \lambda$, so that the work term \textit{is} uniquely identified. \textit{Hence the fact that the expressions obtained for  $S$ and $E$ by this procedure match exactly with the known expressions for spherically symmetric horizons is non-trivial.} In particular, the general expression for energy for arbitrary static spacetimes has a simple geometric expression, and deserves further study.

Finally, we would like to clarify the difference between the first law of thermodynamics as obtained above, and the usual first law of black hole \textit{mechanics}. In the conventional case, one obtains the first law by varying the parameters in a \textit{specific solution} to the field equations. Our result above shows that the field equations governing the \textit{dynamics} of gravity themselves have a thermodynamic structure, and can be uniquely (see the discussion in the preceding paragraph) written in the form of the first law of thermodynamics. In general, the first law we have obtained would be different from the conventional first law of black hole mechanics, although, as described in section \ref{sec:eh-review}, these match in the case of spherical symmetry (the reason can be traced back to the fact that the transverse metric in spherical symmetry does not depend on the parameters in the metric). Because of the specific meaning we have attached to the differentials in the first law, the variations we are dealing with are best looked upon as arising due to \textit{virtual displacement} of the horizon. 
\section{Discussion} \label{sec:discussion}
It has been well known, for quite some time now, that the gravitational field equations near the horizon can be written as a thermodynamic identity; this fact has been demonstrated by a large number of specific examples \cite{gravtherm2, aseem-lovelock}. Our present work can be considered as a formal proof of this intriguing connection between gravitational dynamics and horizon thermodynamics, for \textit{general, static spacetimes}. In the above analysis, we have only demonstrated that the $t-n$ part of the field equations can be so written. However, this is the only relevant part since the horizon is \textit{located} at $t=$ constant, $n=0$; more formally, the observers who perceive the $t=$ constant, $n=0$ surface as a horizon are those moving along the orbits of the Killing vector field generating the horizon. The statement we have proved is, in fact, that the relations
\begin{eqnarray}
E( \bm{\hatn}, \bm{\hatn} ) \;|_H &=& E( \bm{\hatxi}, \bm{\hatxi} ) \;|_H \\
\nonumber \\
E( \bm{\hatxi}, \bm{\hatxi} ) &=& \frac{1}{2} \; T( \bm{\hatxi}, \bm{\hatxi} ) 
\end{eqnarray}
with $\bm{\hatxi} \cdot \bm{\hatxi} = -1 = - \bm{\hatn} \cdot \bm{\hatn}$, represent a thermodynamic identity. There would, of course, exist different class of observers with different horizons, and so long as the background is static, the above statement must be true for all of them. Imposing this then leads to the full field equations, $E_{a b} = (1/2) T_{a b}$. Moreover, since we have related the full gravitational field equations to the thermodynamics of horizons, this indicates an essentially holographic nature of gravitational dynamics, with \textit{classical} symmetries near the horizon, along with the first law of thermodynamics, governing the entire gravitational dynamics \cite{ayan}.
In addition to obtaining the standard field equations, such an approach also leads to a quantisation condition on entropy \cite{entquant} which reduces to quantisation of areas at the lowest order. Indeed, there is a much more formal and general way of obtaining the field equations of Lanczos-Lovelock gravity by taking the thermodynamic interpretation as a starting point, and using normals to null surfaces as the relevant degrees of freedom \cite{paddy-nullvecs}.  

Recently, it was shown in \cite{paddy-ent-density} that for any diffeomorphism invariant theory, the local thermodynamic relation $T \DM S = \DM E$ holds provided one interprets $S$ as a suitable Noether current associated with diffeomorphism invariance, \textit{and defined off-shell}; this is important since one must not use quantities defined on-shell while trying to ``derive" the field equations. The arguments presented in \cite{paddy-ent-density} apply to any diffeomorphism invariant theory, whereas our result in this paper relies heavily on the near-horizon symmetries of the field tensor. Although a direct connection between the two is not immediately apparent (particularly, our expression for $\DM E$ is different, and we have an additional $P \DM V$ term in the first law), it would be interesting to see whether our arguments can be generalised to any diffeomorphism invariant theory, in the light of the results in \cite{paddy-ent-density}.

At a deeper level, these results suggest that it is necessary to abandon the usual picture of treating the
metric as  the fundamental dynamical degrees of freedom of the theory and treat\cite{sakharov} it as
 providing a
coarse grained description of the spacetime at macroscopic scales,
somewhat like the density of a solid ---  which has no meaning at atomic  
scales. The unknown, microscopic degrees of freedom of
spacetime (which should be analogous to the atoms in the case of
solids), should normally play a role only when spacetime is probed at Planck
scales (which would be analogous to the lattice spacing of a solid
\cite{zeropoint}).  So we normally expect the  microscopic structure of spacetime
to manifest itself only at Planck scales or near singularities of the
classical theory. However, in a manner which is not fully understood,
the horizons ---  which block information from certain classes of
observers --- link \cite{magglass} certain aspects of microscopic
physics with the bulk dynamics, just as thermodynamics can provide a
link between statistical mechanics and (zero temperature) dynamics of
a solid. The  reason is probably related to the fact that horizons
lead to infinite redshift, which probes \textit{virtual} high energy
processes; it is, however, difficult to establish this claim in
mathematical terms. This aspect, as to why horizons act as window to microphysics of spacetime, is worth investigating further. 

Finally, we would like to mention two immediate possible extensions of the proof given here: (i) \textit{stationary, non-static spacetimes}, and (ii) \textit{time-dependent spacetimes with horizons}. In the case of stationary, non-static horizons, one needs to have a clear notion of quantities such as \textit{normal pressure} [since the intrinsic horizon geometry is non-trivial]. However, it has been shown that the Einstein equations can indeed be expressed as a thermodynamic identity in the specific case of Kerr-Newman black hole (see the last reference in \cite{gravtherm2}). Combining this with the general analysis in the second reference in \cite{visser1}, one expects an analogous result to exist for the stationary, non-static case as well. For the time-dependent case, determining the near horizon form of the field equations itself would be more involved. Apart from this, there are certain conceptual issues such as the notion of temperature, which must be addressed. However, it must still be possible to operationally define a temperature (at least in some quasi-static sense), and see how far the near horizon symmetries still conspire to give a thermodynamic structure to the field equations.

\section*{ACKNOWLEDGEMENTS} 
The authors would like to thank Aseem Paranjape for comments on the manuscript. DK is supported by a Fellowship from the Council of Scientific \& Industrial Research (CSIR), India.
\appendix
\section{Near-horizon symmetries of the Einstein tensor} \label{app1}
In this appendix, we briefly outline the proof of Eq.~(\ref{near-hor-eins}). We begin with the following expressions for the decomposition of the Riemann tensor for the metric (\ref{static-metric}) [see, for example, section 21.5 and Exercise 21.9 of \cite{mtw}]. 
\begin{eqnarray}
R^{\tt}_{~ \mu \nu \rho} &=& 0 \nonumber \\
R_{\mu \tt \nu \tt} &=& N N_{| \mu \nu} \nonumber \\
R^{\mu}_{~ \nu \rho \sigma} &=& ~^{(3)} R^{\mu}_{~ \nu \rho \sigma} 
\label{rd11}
\end{eqnarray}
and
\begin{eqnarray}
~^{(3)} R_{\small{A B C D}} &=& ~^{(2)} R_{\small{A B C D}} - \l( K_{\small{A C}} K_{\small{B D}} - K_{\small{A D}} K_{\small{B C}} \r) \nonumber \\
~^{(3)} R_{\tn \small{B C D}} &=& K_{\small{A C} :\small{B}} - K_{\small{A B} :\small{C}} 
\label{rd1}
\end{eqnarray}
Here, $|$ and $:$ are the covariant derivatives compatible with the induced metric on \{$t=$ constant\} and \{$t=$ constant, $n=$ constant\} surfaces respectively, and $K_{\small{A B}} = -(1/2) \partial_n \sigma_{\small{A B}}$ is the extrinsic curvature of the \{$t=$ constant, $n=$ constant\} 2-surface as embedded in the \{$t=$ constant\} surface. These expressions lead to  
\begin{eqnarray}
\rnna &=& \sigma^{{\small A} {\small C}} \left[ \partial_n K_{{\small C} {\small B}} + \left( K^2 \right)_{{\small B} {\small C}} \right]
\nonumber \\
\rtta &=& \sigma^{{\small A} {\small C}} \left[ - \frac{N_{:{\small C} {\small B}} - K_{{\small C} {\small B}} \partial_n N}{N} \right]
\label{rd2} 
\end{eqnarray}
Using the relevant Taylor series expansions, we obtain
\begin{eqnarray}
\partial_n \mathrm{tr}K |_{n=0} = - \frac{1}{2} \mathrm{tr}[\sigma_2]
\end{eqnarray}
which gives, correct to $\mathcal{O}(n^2)$,
\begin{eqnarray}
\rnn = - \frac{1}{2} \mathrm{tr}[\sigma_2] = \rtt  
\label{rd3} 
\end{eqnarray}
and
\begin{eqnarray}
R^{{\small A B}}_{~ {\small C D}} = {}^{(2)}R^{{\small A B}}_{~ {\small C D}} + \mathcal{O}(n^2)
\nonumber
\end{eqnarray}
which implies
\begin{eqnarray}
R^{\small{A B}}_{~ \small{A B}} = \; R_{\parallel} + \mathcal{O}(n^2)
\label{rd4}   
\end{eqnarray}

Finally, we use the following general expression for the Einstein tensor (see, for example, \cite{mtw}, section 14.2, pp. 344):
\begin{eqnarray}
G^{\tt}_{\tt} &=& - \left( \rnn + \frac{1}{2} R^{{\small A B}}_{~ {\small A B}} \right) \nonumber
\\ 
G^{\tn}_{\tn} &=& - \left( \rtt + \frac{1}{2} R^{{\small A B}}_{~ {\small A B}} \right)
\label{mtw-einstein-tensor}
\end{eqnarray}
Plugging in the above expressions for the Riemann tensor, and finally taking the limit $n \rightarrow 0$, we immediately obtain Eq.~(\ref{near-hor-eins}). (Note that, with the up-down components, $G^{\hatxi}_{\hatxi} = G^{\tt}_{\tt}$.) One can go further and analyze, in the same way, the remaining components of the Einstein tensor; the final result is \cite{visser1}
\begin{eqnarray}
E^{\hat a}_{\hat b}|_H = \left[ \begin{array}{cc|c}
E_{\perp} &0 &0\\
0& E_{\perp} & 0\\
\hline
0&  0 & {E_{\parallel }}^{\small{\hat A}}_{\small{\hat B}} \vphantom{\bigg{|}} \end{array} 
\right] \;
\label{eom-matrix}
\end{eqnarray}
where $E^{\hat a}_{\hat b}|_H = \l( 16 \pi \r)^{-1} G^{\hat a}_{\hat b}|_H$.
\section{Equivalence of the two expressions in Eqs.\;(\ref{eom-lovelock})} \label{app:LLeom}
We need to prove the equality;
\begin{eqnarray}
\l[ \delta^{i}_{j} \delta^{a_1 b_1 \ldots a_m b_m}_{c_1 d_1 \ldots c_m d_m} - 2 m \; \delta^{i}_{c_1} \delta^{a_1 b_1 \ldots a_m b_m}_{j d_1 \ldots c_m d_m} \r] R^{c_1 d_1}_{~ a_1 b_1} \cdots R^{c_m d_m}_{~ a_m b_m} 
\nonumber \\
= \delta^{i a_1 b_1 \ldots a_m b_m}_{j c_1 d_1 \ldots c_m d_m} R^{c_1 d_1}_{~ a_1 b_1} \cdots R^{c_m d_m}_{~ a_m b_m}
\nonumber \\
\label{eq:alter-tensor}
\end{eqnarray}
This is most easily done by noting that the alternating tensor on the RHS of Eq.\;(\ref{eq:alter-tensor}) can be written as a determinant:
\begin{eqnarray}
\delta^{i a_1 b_1 \ldots a_m b_m}_{j c_1 d_1 \ldots c_m d_m} = {\mathrm{det}} \left[ \begin{array}{c|ccc}
\delta^i_j & \delta^i_{c_1} & \cdots & \delta^i_{d_m} 
\\
\hline
\\
\delta^{a_1}_j &  & & 
\\
\vdots & & \delta^{a_1 b_1 \ldots a_m b_m}_{c_1 d_1 \ldots c_m d_m} &
\\
\delta^{b_m}_j & & &  \vphantom{\bigg{|}} \end{array} 
\right] \;
\end{eqnarray}
The first term on the LHS of Eq.\;(\ref{eq:alter-tensor}) therefore comes from the multiplication of $\delta^i_j$ with the lower right block of the above matrix. The remaining terms in the determinant can be grouped as 
\begin{eqnarray}
&-& \l[ \delta^i_{c_k} \delta^{a_1 b_1 \;\; \cdots \;\; a_m b_m}_{j c_1 d_1 \cdots d_{k-1} d_{k} \cdots c_m d_m} - \delta^i_{d_k} \delta^{a_1 b_1 \;\; \cdots \;\; a_m b_m}_{j c_1 d_1 \cdots c_{k} c_{k+1} \cdots c_m d_m} \r]
\nonumber \\
\label{eq:alter-tensor-1}
\end{eqnarray}
where $1 \le k \le m$. Since this whole determinant is multiplied by the product of curvature tensors, only the piece antisymmetric in the pair $\{c_k, d_k\}$ will be picked up, producing a factor of $2$ for each pair $\{c_k, d_k\}$. Further, each of the $m$ such pairs contribute the same amount due to the symmetries of the alternating tensor and the curvature tensor. This gives another factor of $m$, so that the contribution of the remaining terms in the above determinant becomes equal to $2 m$ times the contribution of any particular term, say the term corresponding to the pair $\{c_1, d_1\}$. Noting the overall minus sign in (\ref{eq:alter-tensor-1}), we obtain the second term on the LHS of Eq.\;(\ref{eq:alter-tensor}), thereby proving the desired result.
\section{Near-horizon symmetries of the Lanczos-Lovelock field equations} \label{app2}
One can do an analysis similar to that outlined in Appendix \ref{app1} and obtain the near-horizon symmetries of the LL field equations. The full result, which we state without proof (the proof involves a bit of combinatorics), turns out to be the same as Eq.~(\ref{eom-matrix}) \cite{note-ll-symm}. However, in this Appendix, we shall only prove the identity $E^{\hat \xi}_{\hat \xi} = E^{\hatn}_{\hatn} ~ |_H$, which is directly relevant for our purpose.

Using the first equality in Eq.~(\ref{eom-lovelock}), it is easy to deduce the form of $E^{\textsl k}_{\textsl k}$ (no summation over $k$). The alternating determinant simplifies upon using $\delta^{\textsl k}_{\textsl k} = 1$, and the fact that none of the other indices can be $k$ due to total anti-symmetry. Therefore, we are left with
\begin{eqnarray}
E^{\textsl k}_{\textsl k} = - \frac{1}{2} \mathcal{L}_m \l\{ \overline{k} \r\}
\end{eqnarray}
where $\mathcal{L}_m \l\{ \overline{k} \r\}$ denotes terms which do not contain $k$ at all. We now specialise to $k=n$. The RHS can be further split depending on the number of occurences of the index $\tt$, as
\begin{eqnarray}
E^\tn_\tn = - \frac{1}{2} \l[ \mathcal{L}_m \l\{ \overline{n}, t \r\} + \mathcal{L}_m \l\{ \overline{n}, 2 t \r\} + \mathcal{L}_m \l\{ \overline{n}, \overline{t} \r\} \r]
\end{eqnarray}
The first set on the RHS contains terms like $R^{\tt {\small D}}_{~\small A B}$ which are identically zero (see first of Eqs.~(\ref{rd11})), while the last set is the same as appears in $E^\tt_\tt$. So we only have to prove that $\mathcal{L}_m \l\{ \overline{n}, 2 t \r\} = \mathcal{L}_m \l\{ \overline{t}, 2 n \r\} $. The set $\mathcal{L}_m \l\{ \overline{n}, 2 t \r\}$ will have two $t$'s appearing either on \textit{different} factors of $R^{a b}_{~c d}$, which would again vanish identically for the same reason as the first set, or it can have the two $t$'s appearing on the \textit{same} factor, which would contribute
\begin{eqnarray}
\mathcal{L}_m \l\{ \overline{n}, 2 t \r\} &=& 4m \times \frac{1}{16 \pi} \frac{1}{2^m} \delta^{\tt {\small B}_1 \ldots {\small A}_m {\small B}_m}_{\tt {\small D}_1 \ldots {\small C}_m {\small D}_m} R^{\tt {\small D}_1}_{~ \tt {\small B}_1} \cdots R^{{\small C}_m {\small D}_m}_{~ {\small A}_m {\small B}_m}
\nonumber \\
&=& 4m \times \frac{1}{16 \pi} \frac{1}{2^m} \delta^{{\small B}_1 \ldots {\small A}_m {\small B}_m}_{{\small D}_1 \ldots {\small C}_m {\small D}_m} R^{\tt {\small D}_1}_{~ \tt {\small B}_1} \cdots R^{{\small C}_m {\small D}_m}_{~ {\small A}_m {\small B}_m}
\nonumber \\
\end{eqnarray}
Using Eqs.~(\ref{rd2}), we see that $\mathcal{L}_m \l\{ \overline{n}, 2 t \r\} = \mathcal{L}_m \l\{ \overline{t}, 2 n \r\} + \mathcal{O}(n^2)$. Therefore, $E^\tn_\tn = E^{\tt}_{\tt} + \mathcal{O}(n^2)$, and the equality holds on the horizon.
%



\begin{thebibliography}{99}
\bibitem{note-ll-symm}
This fact, which generalizes the result of the first reference in \cite{visser1} to LL gravity, depends on the specific way in which the LL lagrangians are constructed; an arbitrary higher derivative theory might not, in general, have such symmetries.
\end{thebibliography}
\end{document}